\documentstyle[prl,aps,twocolumn,epsfig,floats]{revtex}

\begin{document}

\def\eovera{$\left< E/A\right> _{\rm PLF}$}
\def\xeau{$^{129}$Xe+$^{197}$Au} 
\def\xebi{$^{136}$Xe+$^{209}$Bi}
\def\Etlcp{$E_t^{\rm LCP}$} \def\aveEtlcp{$\left< E_t^{\rm
LCP}\right>$} \def\Etimf{$E_t^{\rm IMF}$} \def\aveEtimf{$\left<
E_t^{\rm IMF}\right>$} \def\mimf{$N_{\rm IMF}$} \def\mlcp{$N_{\rm
LCP}$} \def\avemimf{$\left< N_{\rm IMF}\right>$} \def\avemlcp{$\left<
N_{\rm LCP}\right>$} \def\mimfsat{$N_{\rm IMF}^{\rm sat}$}
\def\avemlcpmax{$\left< N_{\rm LCP}\right>_{\rm max}$}
\def\aveEtlcpmax{$\left< E_t^{\rm LCP}\right>_{\rm max}$}
\def\avemimfmax{$\left< N_{\rm IMF}\right>_{\rm max}$}



\title{A statistical interpretation of the correlation between \\
intermediate mass fragment
multiplicity and transverse energy}

\author{L. Phair$^1$, L. Beaulieu$^{1,*}$, L.G. Moretto$^1$,
G.J. Wozniak$^1$, D. R. Bowman$^{2,\dag}$, N. Carlin$^{2,\S}$,
L. Celano$^5$, N. Colonna$^{5}$, J.D. Dinius$^2$, A. Ferrero$^3$,
C.K. Gelbke$^2$, T. Glasmacher$^2$, F. Gramegna$^6$, D.O. Handzy$^2$,
W.C.  Hsi$^2$, M.J.  Huang$^2$, I. Iori$^3$, Y.D. Kim$^{2,\parallel}$,
M.A. Lisa$^{2,\P}$, W.G. Lynch$^2$, G.V. Margagliotti$^4$,
P.F. Mastinu$^{7,**}$, P.M. Milazzo$^{7,**}$, C.P. Montoya$^2$,
A. Moroni$^3$, G.F. Peaslee$^{2,\dag\dag}$, R. Rui$^4$, C. Schwarz$^2$,
M.B. Tsang$^2$, K. Tso$^1$, G. Vannini$^4$, F. Zhu$^{2}$}

\address{$^{1}$Nuclear Science Division, Lawrence Berkeley National
Laboratory, Berkeley, California 94720}

\address{$^2$National Superconducting Cyclotron Laboratory and
Department of Physics and Astronomy, \\ Michigan State University,
East Lansing, MI 48824}

\address{$^3$Istituto Nazionale di Fisica Nucleare and Dipartimento di
 Fisica, Via Celoria 16, 20133 Milano, Italy}

\address{$^4$Dipartimento di Fisica and Istituto Nazionale di Fisica
Nucleare, Via A. Valerio 2, 34127 Trieste, Italy}

\address{$^5$Istituto Nazionale di Fisica Nucleare, Via Amendola 173,
70126 Bari, Italy}

\address{$^6$Istituto Nazionale di Fisica Nucleare, Laboratori
Nazionali di Legnaro, Via Romea 4, 35020 Legnaro, Italy}

\address{$^7$Dipartimento di Fisica and Istituto Nazionale di Fisica
Nucleare, Bologna, Italy}

\date{\today}

\maketitle

\begin{abstract}
Multifragment emission following \xeau ~collisions at 30, 40, 50 and
60 $A$MeV has been studied with multidetector systems covering nearly
4$\pi$ in solid angle. The correlations of both the intermediate mass
fragment and light charged particle multiplicities with the transverse
energy are explored.  A comparison is made with results from a similar
system, \xebi ~at 28 $A$MeV.  The experimental trends are compared to
statistical model predictions.
\end{abstract}


\narrowtext


\section{Introduction}

Nuclear multifragmentation is arguably the most complex nuclear
reaction, involving both collective and internal degrees of freedom to
an extent unmatched even by fission.  As in fission,
multifragmentation is expected to present a mix of statistical and
dynamical features.

A substantial body of evidence has been presented in favor of the
statistical nature of several features such as fragment multiplicities
\cite{Bow91,Des91,Del93,Moretto93,Moretto95,Tso95,Don97,Moretto97,Gro97,Beaulieu98},
charge distributions \cite{Phair95,Dag96}, and angular distributions
\cite{Phair96}. Recently however, evidence has been put forth for the
lack of statistical competition between intermediate mass fragment
(IMF) emission and light charged particle (LCP) emission. More
specifically, it has been shown that for the reaction \xebi ~at 28
$A$MeV: a) LCP emission saturates with increasing number of emitted
IMFs \cite{Tok96}; b) with increasing transverse energy ($E_t$), the
contribution of the LCPs to $E_t$ saturates while that of the IMFs
becomes dominant \cite{Tok97}; c) there is a strong anti-correlation
of the leading fragment kinetic energy with the number of IMFs emitted
\cite{Tok96}. This body of evidence seems to suggest that beyond a
certain amount of energy deposition most, if not all, of the energy
goes into IMF production rather than into LCP emission in a manner
inconsistent with statistical competition.

Given the importance of these results in showing a potential failure
of the statistical picture and a possible novel dynamical mechanism of
IMF production, we have applied the same analysis to a set of
systematic measurements of $^{129}$Xe+$^{197}$Au at several bombarding
energies.  In what follows we report on 1) new experimental data that
confirm the general nature of the observations in \cite{Tok96}; 2) new
experimental data which show trends that are different from those
observed in \cite{Tok97}; 3) the effectiveness of gating on IMF
multiplicity (\mimf ) as an event-selection strategy; and 4) the
reproduction of key results with statistical model calculations.



\section{Experimental setup}

LCP and IMF yields and their correlations with, and contributions to
$E_t$ were determined for the reaction \xeau ~at 30, 40, 50, and 60
$A$MeV.  The experiments were performed at the National
Superconducting Cyclotron Laboratory at Michigan State University
(MSU).  Beams of $^{129}$Xe, at intensities of about $10^7$ particles
per second, irradiated gold targets of approximately 1 mg/cm$^2$.  The
beam was delivered to the 92 inch scattering chamber with a typical
beam spot diameter of 2-3 mm.

For the bombarding energies of 40, 50 and 60 $A$MeV, LCPs and IMFs
emitted at laboratory angles of 16$^{\circ}$-160$^{\circ}$ were
detected using the MSU Miniball \cite{miniball}. As configured for
this experiment, the Miniball consisted of 171 fast plastic (40
$\mu$m)-CsI(2 cm) phoswich detectors, with a solid angle coverage of
approximately 87\% of 4$\pi$. Identification thresholds for $Z$=3,10,
and 18 fragments were $\approx$2, 3, and 4 MeV/nucleon,
respectively. Less energetic charged particles with energies greater
than 1 MeV/nucleon were detected in the fast plastic scintillator
foils, but were not identified by $Z$ value. Isotopic identification
was achieved for hydrogen and helium isotopes with energies less than
75 MeV/nucleon. Energy calibrations were performed using elastically
scattered $^{12}$C beams at forward angles and by using the
punch-through points of the more backward detectors to normalize to
existing data \cite{Kim92}. The energy calibrations are estimated to
be accurate to about 10\% at angles less than 31$^{\circ}$ and to
about 20\% for the more backward angles.

Particles going forward ($\le 16^{\circ}$) were measured with the LBL
forward array \cite{LBLarray}, a high resolution Si-Si(Li)-plastic
scintillator array. Fragments of charge $Z$=1-54 were detected with
high resolution using a 16-element Si(300 $\mu$m)-Si(Li)(5
mm)-plastic(7.6 cm) array \cite{LBLarray} with a geometrical
efficiency of $\approx$64\%. Where counting statistics allowed,
individual atomic numbers were resolved for $Z$=1-54. Representative
detection thresholds of $Z$=2, 8, 20 and 54 fragments were
approximately 6, 13, 21, and 27 MeV/nucleon, respectively. Energy
calibrations were obtained by directing 18 different beams ranging
from $Z$=1 to 54 into each of the 16 detector elements. The energy
calibration of each detector was accurate to better than 1\%, and
position resolutions of $\pm$1.5 mm were obtained.

The complete detector system for these higher energies (LBL array +
Miniball) subtended angles from 2$^{\circ}$-160$^{\circ}$ and had a
geometric acceptance $\approx$88\% of 4$\pi$. As a precaution against
secondary electrons, detectors at angles larger than 100$^{\circ}$
were covered with Pb-Sn foils of thickness 5.05 mg/cm$^2$ (this
increased the detection thresholds for these backward detectors). Both
the Miniball and forward array were cooled and temperature stabilized.

For the 30 $A$MeV data set, the forward going particles ($\theta
=8^{\circ}- 23^{\circ}$) were measured by the MULTICS array
\cite{multics}, a high resolution gas-Si-Si(Li)-CsI array. Detection
thresholds were approximately 2.5 MeV/nucleon for all fragments
($Z$=1-54), and the resolution in $Z$ was much better than 1 unit for
$Z<$30. Energy calibrations were performed by directing 18 separate
beams into each of the 36 telescopes. The calibration beams had
energies of $E/A$=30 and 70 MeV, and ranged in mass from $^{12}$C to
$^{129}$Xe. An energy resolution of better than 2\% was
obtained. Position calibrations of the Si elements of the MULTICS
array were performed with the procedure of ref.~\cite{pos_calib}. The
angular resolution was estimated to be $\approx$
0.2$^{\circ}$. Charged particles emitted beyond 23$^{\circ}$ were
detected with the Miniball in a setup similar to the higher bombarding
energies described above. The complete detector system covered
approximately 87\% of 4$\pi$.

Data were taken
under two trigger conditions: at least two Miniball elements triggered
or at least one IMF observed in the relevant forward array.

Further details of the experimental setups can be found in
refs.~\cite{Bow92,multics_setup}.


\section{Comparison with previous results}

\begin{figure}
\centerline{\psfig{file=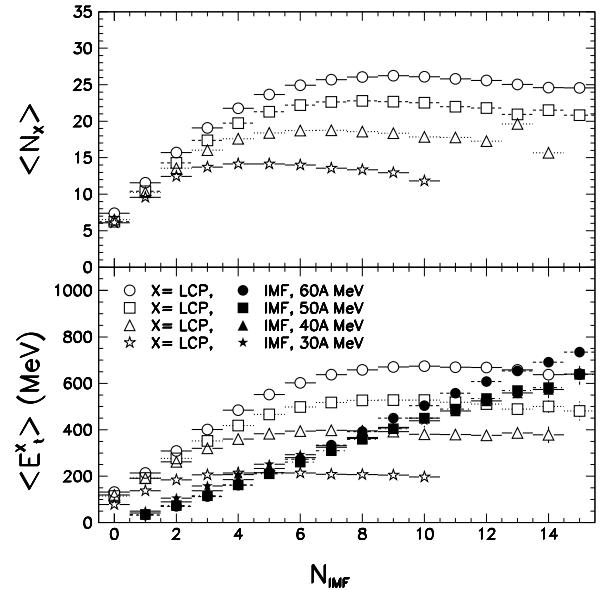,height=8.0cm,angle=0}}
\caption{The average LCP multiplicity (top panel), average transverse
energy of IMFs (solid symbols), and average $E_t$ of LCPs (open
symbols, bottom panel) are plotted as a function of IMF multiplicity
for the reaction $^{129}$Xe+$^{197}$Au.  at bombarding energies
between 30 and 60 $A$MeV.  }
\label{fig:for_n}
\end{figure}

\begin{table}

\caption{Values are given for the approximate \mimf ~saturation value
(along with the upper limit of the integrated cross section in
percent), the average LCP multiplicity and average \Etlcp ~in the
saturation region of Fig.~\protect\ref{fig:for_n}, and the maximum
average IMF multiplicity for the top 5\% of the $E_t$ selected events,
for the reaction \protect\xeau .}

\begin{tabular}{|c|c|c|c||c|}
\hline $E_{\rm beam}/A$ & \mimfsat & \avemlcpmax & \aveEtlcpmax &
\avemimfmax\\ \hline\hline 30 MeV & 5 (6.5\%) & 13.9 & 220 MeV & 4.6\\
40 MeV & 7 (5.0\%) & 18.9 & 400 MeV & 6.0\\ 50 MeV & 8 (4.3\%) & 23.1
& 530 MeV & 6.9 \\ 60 MeV & 8 (5.9\%) & 26.1 & 660 MeV & 7.4 \\
\end{tabular}
\label{table:xeau}
\end{table}

Following the procedure outlined in \cite{Tok96}, the average LCP
yields were determined as a function of \mimf ~(which serves as a
rough measure of impact parameter or energy deposition).
Fig.~\ref{fig:for_n} shows an example of such an analysis for the
reaction \xeau ~at bombarding energies between 30 and 60 $A$MeV.  The
average LCP multiplicity (\avemlcp ) does indeed saturate with
increasing \mimf , as observed in \cite{Tok96}.  However, the value to
which \avemlcp ~saturates (\avemlcpmax ) rises with increasing
bombarding energy and is listed in Table \ref{table:xeau}.  The IMF
multiplicity at which the saturation occurs is approximately 4-5 at 30
$A$MeV and rises with increasing bombarding energy to a value of 8-9
at 60 $A$MeV.  

The average LCP contribution to $E_t$ (\aveEtlcp)
saturates in a bombarding energy dependent fashion as well (see
\aveEtlcpmax ~in Table \ref{table:xeau} and open symbols of
Fig.~\ref{fig:for_n}, bottom panel).  In contrast, the average IMF
contribution to $E_t$ (\aveEtimf ) rises linearly with increasing IMF
multiplicity.  The significance of the bombarding energy dependence of
these observations will be discussed in the next section.

\begin{figure}
\centerline{\psfig{file=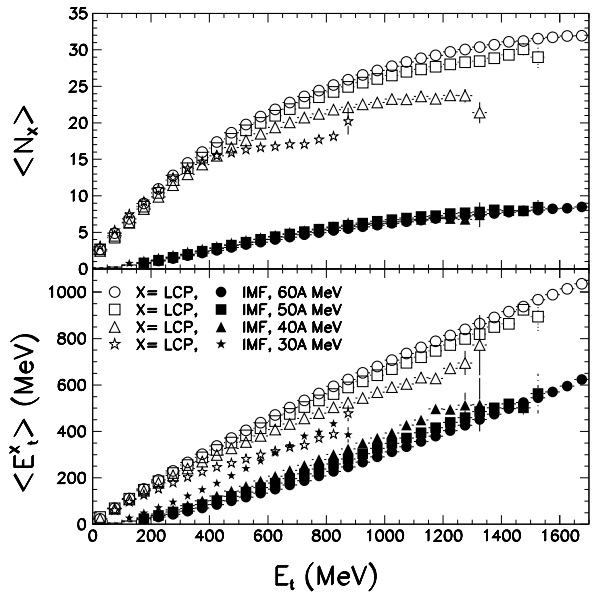,height=8.0cm,angle=0}}
\caption{The average IMF multiplicity (solid symbols, top panel),
average LCP multiplicity (open symbols, top panel), average transverse
energy of IMFs (solid symbols, bottom panel), and average transverse
energy of LCPs (open symbols, bottom panel) are plotted as a function
of $E_t$.}
\label{fig:for_Et}
\end{figure}

\begin{figure} 
\centerline{\psfig{file=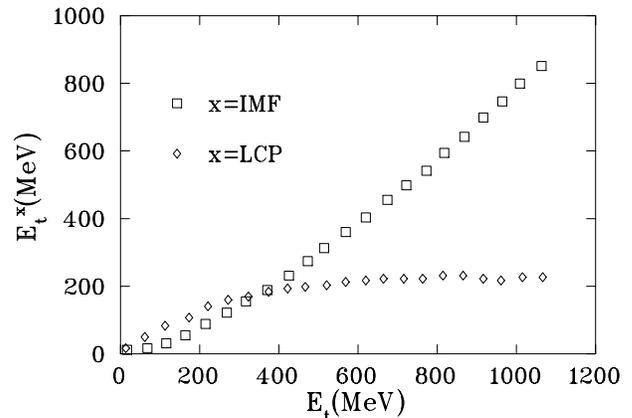,angle=90,height=5.5cm}}
\caption{The average transverse energies of IMFs (squares) and of LCPs
(diamonds) are plotted as a function of $E_t$ for the reaction
\protect\xebi ~at 28 $A$MeV (taken from ref.~\protect\cite{Tok97}).}
\label{fig:early}
\end{figure}

We now explore the dependence of these same variables on $E_t$.
According to the procedure outlined in \cite{Tok97}, the average yields of
multiplicity and transverse energy for both IMFs and LCPs were determined as a
function of $E_t$ (which serves as a measure of impact parameter or energy
deposition \cite{Cav90,Phair92,Phair93}).  
In Fig.~\ref{fig:for_Et} are plotted \avemimf , \avemlcp , \aveEtimf ,
and \aveEtlcp ~as a function of $E_t$ for bombarding energies between
30 and 60 $A$MeV.  All the observables rise with increasing $E_t$, in
disagreement with the observations in \cite{Tok97}.  In \cite{Tok97},
the value of \aveEtlcp ~is observed to saturate to a relatively small
value compared to \aveEtimf (see 
Fig.~\ref{fig:early}), which is at variance with the observations in
Fig.~\ref{fig:for_Et}. The origin of this disagreement will be
discussed in the next section.

\begin{figure}
\centerline{\psfig{file=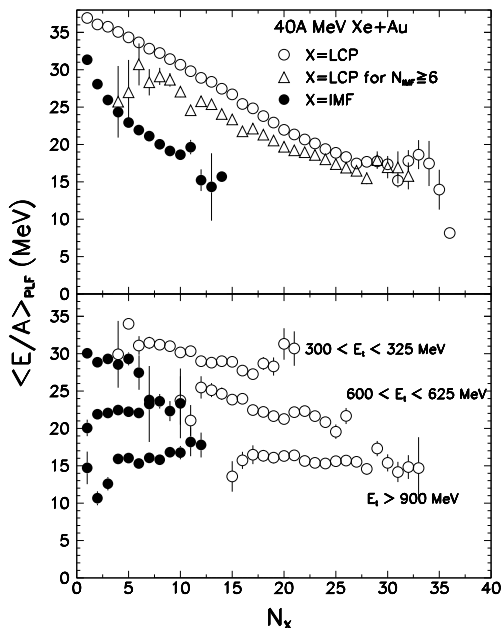,angle=0,height=8.5cm}}
\caption{Top panel: the average kinetic energy per nucleon of the
projectile-like fragment is plotted as a function of \mimf ~(solid
circles) and \mlcp ~(open symbols). Bottom panel: Same as top panel
but selected from events within the indicated range of $E_t$.}
\label{fig:E_over_A}
\end{figure}

Lastly, according to the procedure in \cite{Tok96}, the average
kinetic energy of the projectile-like fragment (\eovera , defined as
the heaviest forward-moving particle in an event, with $Z_{\rm PLF}\ge
10$ and $\theta\le 23^\circ$) has been determined as a function of
\mimf , an example of which is given in Fig.~\ref{fig:E_over_A}.
Here, we confirm the observation in \cite{Tok96}.  For increasing
\mimf , the energy per nucleon of the leading fragment decreases
continuously.

The three aforementioned observations have been used to suggest that,
above a certain excitation energy, the IMFs get the lion's share of
the energy while the LCPs lose their capability to compete
\cite{Tok96,Tok97}.  In the following section, we explore each of these
observations and suggest possible alternative explanations.

\section{Interpretation}

We begin with the saturation of \avemlcp ~and \aveEtlcp ~as opposed to
the continuous rise of \aveEtimf ~observed in Fig.~\ref{fig:for_n}.
\aveEtimf ~rises linearly since
\begin{equation}
\left< E_t^{\rm IMF}\right> = \left<\sum _{i=1}^{N_{\rm IMF}} E_i\sin
^2\theta _i\right>\approx N_{\rm IMF} \left<\epsilon _t^{\rm
IMF}\right>,
\end{equation}
where $\left<\epsilon _t^{\rm IMF}\right>$ is the average transverse
energy of an IMF. Thus, the reason for the continuous rise of
\aveEtimf ~can be understood quite simply.  But what is the reason for
the saturation of \aveEtlcp ~and \avemlcp?  We believe that the values
of \mimf ~where \avemlcp ~and \aveEtlcp ~saturate represent the tails
of the IMF multiplicity distribution which are determined by the most
central collisions.

\begin{figure}
\centerline{\psfig{file=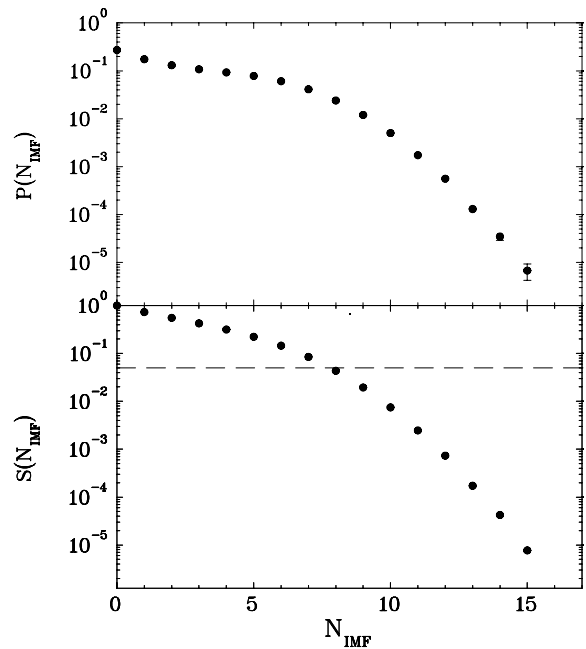,angle=90,height=8.5cm}}
\vspace{0.2cm}
\caption{Top panel: Probability to emit \protect\mimf ~from the
reaction \protect\xeau ~at 50 $A$MeV. Bottom panel: Integrated
probability to emit \protect\mimf ~or more IMFs.}
\label{fig:raw_data}
\end{figure}

For example, the values of 
IMF multiplicity 
~at which the observables in Fig.~\ref{fig:for_n} saturate (\mimfsat )
can be understood in terms of an impact parameter scale.  Consider the
probability $P$ of emitting \mimf ~and its integrated yield
\begin{equation}
S(N_{\rm IMF})=\sum _{i=N_{\rm IMF}}^{\infty}P(i)
\end{equation}
as shown in Fig.~\ref{fig:raw_data} for the reaction \xeau ~at 50
$A$MeV. Average impact parameter scales, as they are commonly
employed, are proportional to $\sqrt S$ \cite{Cav90}.  Note that the
multiplicities at which saturation occurs represent roughly 5\% of the
total integrated cross section (dashed line in the bottom panel of
Fig.~\ref{fig:raw_data}). The \mimf ~value \mimfsat ~for which
$S\approx 0.05$ is listed in Table \ref{table:xeau} for each of the
different bombarding energies. \mimfsat ~tracks rather well the
maximum average 
IMF multiplicity
(\avemimfmax ) measured for the most central collisions (top 5\% of
events) based upon the $E_t$ scale.

The above observations demonstrate that large IMF multiplicities
(\mimf $>$\avemimfmax ) have small probabilities and represent the
extreme tails of events associated with the most central
collisions. In other words, events with increasing values of \mimf ~in
the saturation region of Fig.~\ref{fig:for_n} do not come from
increasingly more central collisions where more energy has been
dissipated. Thus, \mimf ~is
useful as a global event selector over only a very limited range.

\begin{figure}
\centerline{\psfig{file=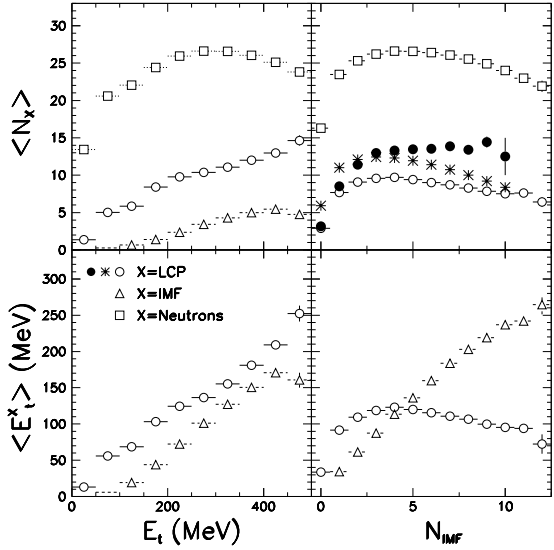,angle=0,height=7.5cm}}
\caption{Statistical model predictions from SMM (open symbols), percolation
(solid symbols), and the simple model described in the text (crossed symbols). 
Upper left: the predicted 
average LCP and IMF multiplicities are plotted
as a function of 
$E_t$ for the decay of an ensemble of gold nuclei with excitation
energies between 0.5-6.0 $A$MeV.  Upper right: the average LCP and
neutron multiplicities are plotted as a function of \protect\mimf .
Lower left: the average $E_t$ of the LCPs and IMFs as a function of
$E_t$
is shown.  Lower right: the average $E_t$ of the LCPs and IMFs is
shown as a function of
\protect\mimf .}
\label{fig:stat_models}
\end{figure}

Consequently, it is expected that statistical models should exhibit
similar trends as those observed in Fig.~\ref{fig:for_n}.  Examples of
such predictions are shown in Fig.~\ref{fig:stat_models} for the
statistical multifragmentation model SMM (open symbols)
\cite{smm_paper} and for percolation (solid symbols) \cite{Bau88}.  In
both models an excitation energy ($E$) distribution was used to mimic
an impact parameter ($b$) weighting.  Assuming that $b$=0 events give
rise to the maximum excitation energy ($E_{\rm max}$), we have chosen
the number of events at a given $E$ proportional to $E_{\rm max}-E$.
The ``excitation energy'' for the percolation calculation is
essentially represented by the number of broken bonds and is
calculated as per ref.~\cite{Bau88}.

Both calculations show a saturation of \avemlcp ~when plotted as a
function of 
IMF
multiplicity. 
This behavior can be understood in terms of a
simple model.  Consider the statistical emission of two particle
types with barriers $B_1$ and $B_2$ (and $B_2>B_1$). Assume the
emission probabilities are $p_i\propto \exp{[-B_i/T]}$ ($i=1,2$) with
$p_1+p_2=1$. With the temperature $T$ characterized in terms of the
total multiplicity $n_{\rm tot}=n_1+n_2=\alpha T$, and ignoring mass
conservation, the solution for $\left< n_1\right>$ as a function of
$n_2$ can be calculated for a distribution of excitation energies like
that described above.  The solution of this
model
is shown by the asterisk symbols in the top right panel of
Fig.~\ref{fig:stat_models} for $B_1$=8, $B_2$=24, $T_{\rm max}$=10 and
$\alpha=2$ (and \mimf =$n_2$, \mlcp=$n_1$). This saturation is
qualitatively 
similar to that of the other statistical models listed in
Fig.~\ref{fig:stat_models} and to the behavior observed in
Fig.~\ref{fig:for_n}. Furthermore, the saturation value of \avemlcp ,
as well as the value of \mimf ~at which saturation occurs, both depend
on the maximum energy used in the calculation. Consequently, for
statistical emission one expects (and observes in
Fig.~\ref{fig:for_n}) a bombarding energy dependence of the saturation
which reflects the total energy available to the decaying system.
These behaviors are generic features that are present in any
statistical model \cite{Phair97}.

For completeness, the IMF and LCP yields from the SMM calculations are
plotted as a function of $E_t$ as well in Fig.~\ref{fig:stat_models}
(left panels).  There is no saturation of \aveEtlcp ~with increasing
$E_t$ as was observed in \cite{Tok97}.  Instead, this calculation
shows qualitatively the same trends as experimentally observed in
Fig.~\ref{fig:for_Et}.

\begin{figure}
\centerline{\psfig{file=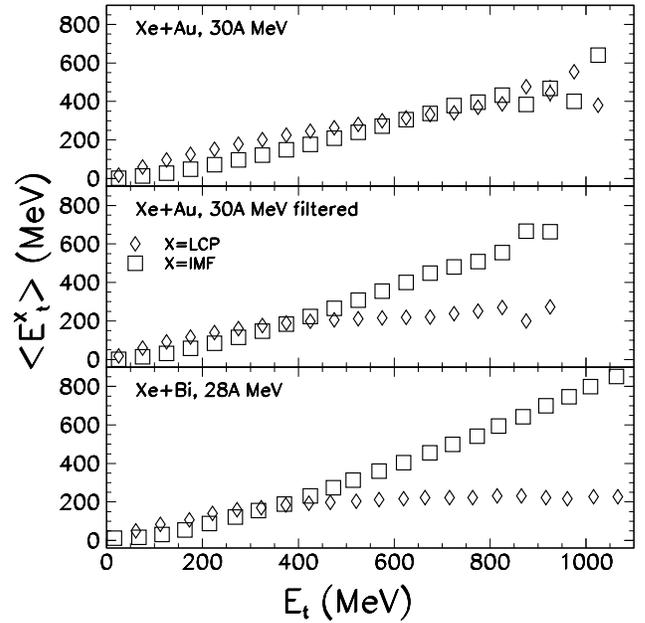,height=8.5cm,angle=0}}
\caption{The average transverse energies of IMFs (squares) and of LCPs
(diamonds) are plotted as a function of $E_t$
for the reactions $^{129}$Xe+$^{197}$Au at 30 $A$MeV (top panel),
\protect\xeau ~again but filtered with the upper energy thresholds of
the dwarf array detector \protect\cite{dwarf} (middle panel), and
$^{136}$Xe+$^{209}$Bi at 28 $A$MeV (bottom panel, taken from
ref.~\protect\cite{Tok97}).}
\label{fig:filter}
\end{figure}

What then causes the (unconfirmed) saturation of \aveEtlcp ~observed
in \xebi \cite{Tok97} (bottom panel of Fig.~\ref{fig:filter})?  We
believe that the saturation observed in
\xebi ~
is likely due to the limited dynamic range of the detectors used.  The
charged particle yields from the \xebi ~reaction were measured with
the dwarf array \cite{dwarf} whose thin CsI crystals (thickness of 4
mm for polar angle $\theta =55-168^{\circ}$, 8mm for
$\theta=32-55^{\circ}$ and 20 mm for $\theta =4-32^{\circ}$) are
unable to stop energetic LCPs.  For example, protons punch through 4
mm of CsI at an energy of 30 MeV. Consequently, their contribution to
$E_t$ could be significantly underestimated.

An example of the distortions that would be caused by the detector
response
of the dwarf array on the
similar \xeau ~reaction at 30 $A$MeV is given in
Fig.~\ref{fig:filter}.  In the top panel is plotted \aveEtlcp ~and
\aveEtimf ~as a function of $E_t$ as measured
by the MULTICS/Miniball collaboration. The thicknesses of the CsI
crystals from these detectors range from 20 to 40 mm.  Protons punch
through 20 mm of CsI with an energy of 76 MeV. In the middle panel of
Fig.~\ref{fig:filter}, the \xeau ~data have been ``filtered'' using
the dwarf array high energy cutoffs which remove high energy particles
from $E_t$. After filtering,
the two prominent features observed in the \xebi ~data set
\cite{Tok97} (bottom panel of Fig.~\ref{fig:filter}) then appear in
the filtered data.  Namely, \aveEtlcp ~saturates to a small value and
\aveEtimf ~becomes the ``apparent'' dominant carrier of $E_t$.
These two features are likely to be instrumental in origin and
therefore do not warrant a physical interpretation.

Last of all, we come to the behavior of the average kinetic energy of
the projectile-like fragment \eovera ~
as a function of \mimf , an example of which is given in
Fig.~\ref{fig:E_over_A} for \xeau ~at 40 $A$MeV (solid circles). From
the decrease of \eovera ~with \mimf , it was concluded that kinetic
energy of the PLF is expended for the production of IMFs
\cite{Tok96}. It was also argued that for increasing 
IMF
multiplicity, 
the saturation of \avemlcp ~represents a critical excitation energy
value beyond which no further amount of relative kinetic energy
between the PLF and TLF is converted into heat. In other words, the
IMFs no longer compete with the LCPs for the available energy -- they
get it all.

One can test the consistency of this explanation by studying the same
observable, $\left< E/A\right> _{\rm PLF}$, but now as a function of
\mlcp ~(open symbols, top panel of Fig.~\ref{fig:E_over_A}).
We observe the
same dependence as that of the IMFs -- a monotonic decrease of $\left<
E/A\right> _{\rm PLF}$ with increasing \mlcp ~which reaches a value of
$\approx$17 MeV at the largest multiplicities. This behavior persists
whether we restrict ourselves to the saturation region (\mimf $\ge 6$,
triangles) or not (open circles).
The similar behavior of \eovera ~with respect to \mimf ~and \mlcp
~indicates that the LCPs do compete with the IMFs for the available
energy.

This can be seen more clearly by pre-selecting events with a better
global observable, $E_t$
\cite{Phair92,Phair93,Llo95}, as done
in the bottom panel of Fig.~\ref{fig:E_over_A}.
Once a window of $E_t$ is selected, a corresponding value of \eovera
~is also determined, and there is no longer any strong dependence of
\eovera ~on \mimf ~or \mlcp . In fact, the resulting \mimf ~and \mlcp
~selections both give the {\em same} value of \eovera , consistent
with a scenario
where both species compete for the same available energy.

\section{Conclusions}

In summary, we have made a systematic study of LCP and IMF observables
as a function of IMF multiplicity and transverse energy for the
reaction \xeau ~at bombarding energies between 30 and 60 $A$MeV.

We observe that \avemlcp ~and \aveEtlcp ~saturate as a function of
\mimf ~in a bombarding energy dependent way.  These saturations are
predicted by statistical models and are fundamental features of
statistical decay \cite{Phair97}.  A bombarding energy dependence of
\avemlcp , \aveEtlcp , and \mimfsat ~is expected (and experimentally
observed) within the framework of statistical decay.

In addition, it has been demonstrated in a model independent fashion
that the LCPs compete with the IMFs for the available energy.  By
using $E_t$, a more sensitive event selection is obtained. The
analysis also demonstrates the limited usefulness of event
classification using only \mimf .

We do not observe a saturation of \aveEtlcp as a function of $E_t$ at
any bombarding energy.  The saturation of \aveEtlcp ~as a function of
$E_t$ observed in ref.~\cite{Tok97} is likely due to instrumental
distortions. We can account for this saturation by filtering the
present measurements of \xeau ~with the experimental thresholds
present in
refs.~\cite{Tok96,Tok97}.  The resulting distortions to the data are
large and induce qualitative changes in the trends of the data,
causing an unphysical saturation of \aveEtlcp . Therefore, the
observations listed in \cite{Tok96,Tok97} do not demonstrate any
measurable failure of statistical models that would justify invoking
dynamical IMF production by default. While the IMFs may indeed be
produced dynamically, the observations listed in
refs.~\cite{Tok96,Tok97} do not provide evidence for such a
conclusion.

Acknowledgments

This work was supported by the 
Nuclear Physics Division of the US Department of Energy, 
under contract DE-AC03-76SF00098, and by the National Science Foundation under
Grants No. PHY-8913815, No. PHY-90117077, and No. PHY-9214992. One of us (L.B)
acknowledges a fellowship from the National Sciences and Engineering
Research Council (NSERC), Canada, and another (A.F.) acknowledges economic
support from the Fundaci\'on J.B. Sauberan, Argentina.

Present addresses:\\
$^{*}$ Indiana University Cyclotron Facility, 2401 Milo B. Sampson
Ln, Bloomington, IN 47408\\ 
$^{\dag}$ Washington Aerial Measurements
Operations, Bechtel Nevada, P.O. Box 380, Suitland, MD 20752\\
$^{\S}$ Instituto de Fisica, Universidade de Sao Paulo, C.P. 66318,
CEP 05389-970, Sao Paulo, Brazil\\ 
$^{\parallel}$ Physics Department, Seoul
National University, Seoul, 151-742, Korea.\\ 
$^{\P}$ Physics
Department, Ohio State University, Columbus, OH 43210\\ 
$^{**}$Dipartimento di Fisica and Istituto Nazionale di Fisica Nucleare, 
Via A. Valerio 2, 34127 Trieste, Italy\\ 
$^{\dag\dag}$Physics Department, Hope
College, Holland, MI 49423\\

\end{document}